\documentclass[a4paper, amsfonts, amssymb, amsmath, reprint, showkeys, nofootinbib, twoside,longbibliography, aps]{revtex4-1}
\usepackage[english]{babel}
\usepackage[utf8]{inputenc}
\usepackage[detect-all]{siunitx}
\usepackage{comment}
\usepackage[normalem]{ulem} 
\usepackage[colorinlistoftodos, color=green!40, prependcaption]{todonotes}
\usepackage{amsthm}
\usepackage{mathtools}
\usepackage{xcolor}
\usepackage{graphicx}
\usepackage[left=23mm,right=13mm,top=35mm,columnsep=15pt]{geometry} 
\usepackage{adjustbox}
\usepackage{placeins}
\usepackage[T1]{fontenc}
\usepackage{lipsum}
\usepackage{csquotes}
\usepackage[pdftex, pdftitle={Article}, pdfauthor={Author}]{hyperref} 

\usepackage{soul,xcolor}

\begin{document}

\title{Platicon purification in self-injection locking regime via Gain Switching}
\author{Chengcong~Li\textsuperscript{1,2}}
\author{Tatiana~S.~Tebeneva\textsuperscript{1}}
\author{Valery~E.~Lobanov\textsuperscript{1}}
\author{Junqiu~Liu\textsuperscript{3,4}}
\author{Dmitry~A.~Chermoshentsev\textsuperscript{1,5}}
\author{Igor~A.~Bilenko\textsuperscript{1,2}}
\author{Artem~E.~Shitikov\textsuperscript{1,*}} 

\affiliation{\textsuperscript{1}Russian Quantum Center, 143026 Skolkovo, Russia}
\affiliation{\textsuperscript{2}Faculty of Physics, Lomonosov Moscow State University, 119991 Moscow, Russia}
\affiliation{\textsuperscript{3}International Quantum Academy, 518048, Shenzhen, China}
\affiliation{\textsuperscript{4}Hefei National Laboratory, University of Science and Technology of China, 230088, Hefei, China}
\affiliation{\textsuperscript{5}Moscow Institute of Physics and Technology, Dolgoprudny, Moscow Region 141701, Russia}
\affiliation{\textsuperscript{*} a.shitikov@rqc.ru}



\begin{abstract}

Integrated photonic devices leveraging optical frequency microcombs have emerged as essential instruments for modern photonic systems, prized for their chip-scale footprint, high energy efficiency, and inherent stability. This work introduces a novel approach to microwave photonic oscillator based on generation of a Kerr platicon microcomb by a gain-switched self-injection-locked distributed feedback (DFB) laser diode.
The system leverages direct modulation of the current of the laser diode, generating optical sidebands around the pump line. The beatnote between the Kerr comb lines and sidebands of the modulated pump at a high-speed photodetector allows to obtain tunable, low-noise microwave signals at frequencies close to the platicon repetition rate, which is many times higher than used for gain switching. We experimentally demonstrate continuous microwave signal frequency tuning from 100 MHz to 3.6 GHz by sweeping the modulation frequency, thereby shifting the generated sidebands relative to a platicon beatnote. 
We match third harmonics of the modulation frequency of the gain-switch signal with the free spectral range of the microresonator simultaneously generating the platicon microcomb. That leads to significant spectral purification of the platicon beatnote decreasing its phase noise by more than 30 dB. 
The novel architecture proposed offers a robust and versatile method for microwave synthesis, presenting significant potential for the advancement of integrated microwave photonic systems and their applications in communications, sensing, and metrology.

\end{abstract}
\maketitle

\section{Introduction}

Microwave photonics plays an essential role in numerous scientific and industrial applications \cite{marpaung2019integrated, yao2022microwave}. In particular, integrated silicon photonics dominates microwave synthesis efforts due to its scalability and compatibility with existing manufacturing processes \cite{tao2025silicon}.
One of the most promising methods for generating pure microwave signals spanning several gigahertz to terahertz frequencies is to leverage soliton microcombs produced in high-quality-factor (high-Q) optical microresonators \cite{jin2021hertz, zhang2019terahertz, sun2023applications, kuse2022low}. In soliton microcomb all spectral components are equidistant and mutually coherent, so at the detector it provides the signal at the repetition rate of the soliton with extremely low phase noise \cite{herr2014temporal}. The generation of solitons for microwave synthesis requires laser with low phase noise.
The self-injection locking (SIL) technique has attracted significant attention as a simple and robust method for laser stabilization \cite{kondratiev2023recent}, enabling the straightforward generation of dissipative Kerr solitons or platicons in microresonators \cite{raja2019electrically, jin2021hertz, shen2020integrated}. It offers exceptional properties such as stability, energy conversion efficiency, and compactness of the device \cite{ye2022integrated, gil2025high}.
The combination of microwave synthesis through the SIL frequency comb with additional electrical feedbacks provides leading results among the field \cite{nakamura2026low, kudelin2024photonic}.
During the past two decades, the rapid advancement of integrated photonic technologies has spurred the design and fabrication of chip-scale tunable microwave source platforms, such as silicon \cite{zhang2018silicon,zhang2023hybrid}, silicon nitride \cite{liu2022tunable,qiu2025large,zang2026wide}, indium phosphide \cite{tang2018integrated}, and lithium niobate \cite{he2023high,ma2024ka,ma2024widely,tao2025ultrabroadband}, which have demonstrated significant potential for practical applications. Silicon nitride has attracted special attention due to its low losses and diverse possibilities. Thin Si$_3$N$_4$ microresonators are characterized by normal group velocity dispersion (GVD). At normal GVD, coherent frequency combs exist in the form of platicons \cite{Lobanov:15}. A beatnote of a platicon microcomb obtained via the SIL technique provides an ultra-narrow-linewidth microwave signal with low single-sideband (SSB) phase noise \cite{xue2015mode, lihachev2022platicon, sun2025chip}. However, the frequency adjustment of the Kerr microcomb microwave signal is inherently limited and can be attributed by manipulating the microresonator size with the heater or piezo actuator \cite{Liu2020, lihachev2022low}.

\begin{figure*}[hbtp!]
\centering
\includegraphics[width=\linewidth]{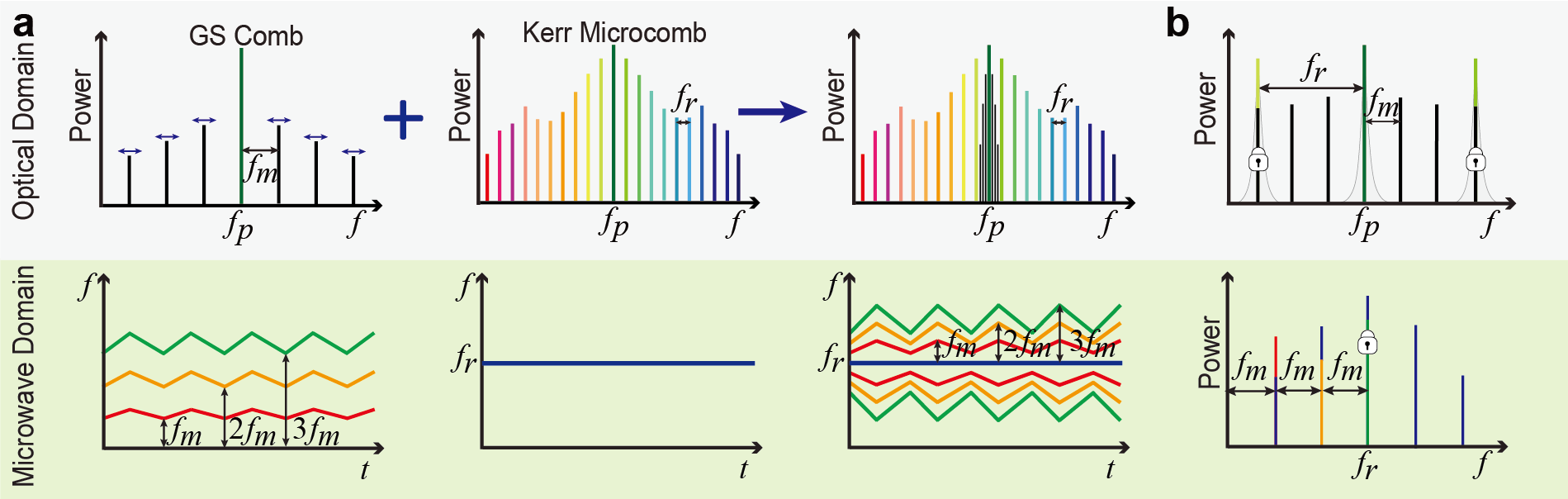}
\caption{(a) The principle of the agile tunable microwave generator based on the gain-switched self-injection-locked microcomb. The GS comb provides one or several harmonics around pump frequency $f_p$ at modulation frequency $f_m$ which is flexible (presented with triangular wave). The Kerr microcomb provides a beatnote at repetition rate $f_r$ and its multiples. Involving both Kerr and GS microcombs the beatnote at repetition rate of the Kerr comb obtains tunable highly coherent sidebands at modulation frequency offset. (b) The concept of the spectral purification of the platicon repetition rate through GS components at the FSR frequency. We use third harmonics GS modulation frequency which exactly matches the FSR of the microresonator for increasing of the long-term stability of the platicon repetition rate. 
}
\label{fig:Concept}
\end{figure*}

Here we combine SIL with the gain-switch (GS) regime of a laser diode. 
Gain switching opens up interesting possibilities for signals in the optical domain by influencing the laser diode with easily tunable and controllable microwave signals \cite{lau1988gain}.
GS laser diodes provide a simple and efficient method for generating optical frequency combs by manipulating the diode laser operation current \cite{anandarajah2011generation, weng2021gain}. As a result, GS lasers are commonly used in spectroscopy \cite{jerez2016dual}, frequency stabilization \cite{liekhus2012injection}, and frequency comb generation \cite{quirce2020nonlinear, zhu2016novel, quevedo2020gain}. At the same time, the GS laser requires additional stabilization, which is conventionally achieved through injection locking \cite{quevedo2020gain, liu2019optical}. The combination of SIL with a GS laser offers a unique advantage: the generation of ultra-narrow-linewidth frequency comb with tunable spacing across different spectral bands \cite{shitikov2021self, shao2022gain, shitikov2023red}. To overcome the limitations of standalone Kerr microcomb beatnote which cannot be tuned and GS comb which requires additional stabilization, we propose to combine these two approaches. Moreover, we demonstrate that the Kerr microcomb beatnotes can be additionally purified via gain-switching.

\section{Results}

\subsection{Concept}

The principle of the experiment is illustrated in Fig.~\ref{fig:Concept}(a). First, the gain-switching regime of the pump DFB laser is activated via the bias-tee circuit, with the modulation frequency $f_m$ determined by the microwave generator. The DFB laser is then self-injection locked to a mode of the normal-dispersion Si$_3$N$_4$ microresonator (MRR), generating a platicon microcomb \cite{sun2025chip}. The output of the fast photodetector contains the beatnotes of all optical signals. Kerr comb components produce a microwave signal at the MRR’s free spectral range (FSR) $f_r$, while the beatnote between the first Kerr comb sidebands and the GS comb sidebands yields frequencies of $f_r\pm f_m$, $f_r\pm 2f_m$ and other higher-order sideband signals. 
The manipulation of the gain-switching modulation frequency $f_m$ provides a continuous, versatile, and precise tuning of the sideband frequency.

Since the GS laser contains several sideband harmonics at frequency intervals $n\cdot f_m$, one may choose the frequency value of $n\cdot f_m$ that exactly matches the repetition rate of the microcomb $f_r$ to lock the beatnote of the microcomb generated in the microresonator. This locking principle is illustrated in Fig.~\ref{fig:Concept}(b). The GS sidebands reach the first sidebands of the microcomb and injection lock the platicon repetition rate. A similar approach was implemented in works \cite{weng2019spectral, liu2020photonic} where through phase modulation at the FSR frequency the beatnote of soliton was additionally disciplined.


\begin{figure}[hbtp!]
\centering
\includegraphics[width=\linewidth]{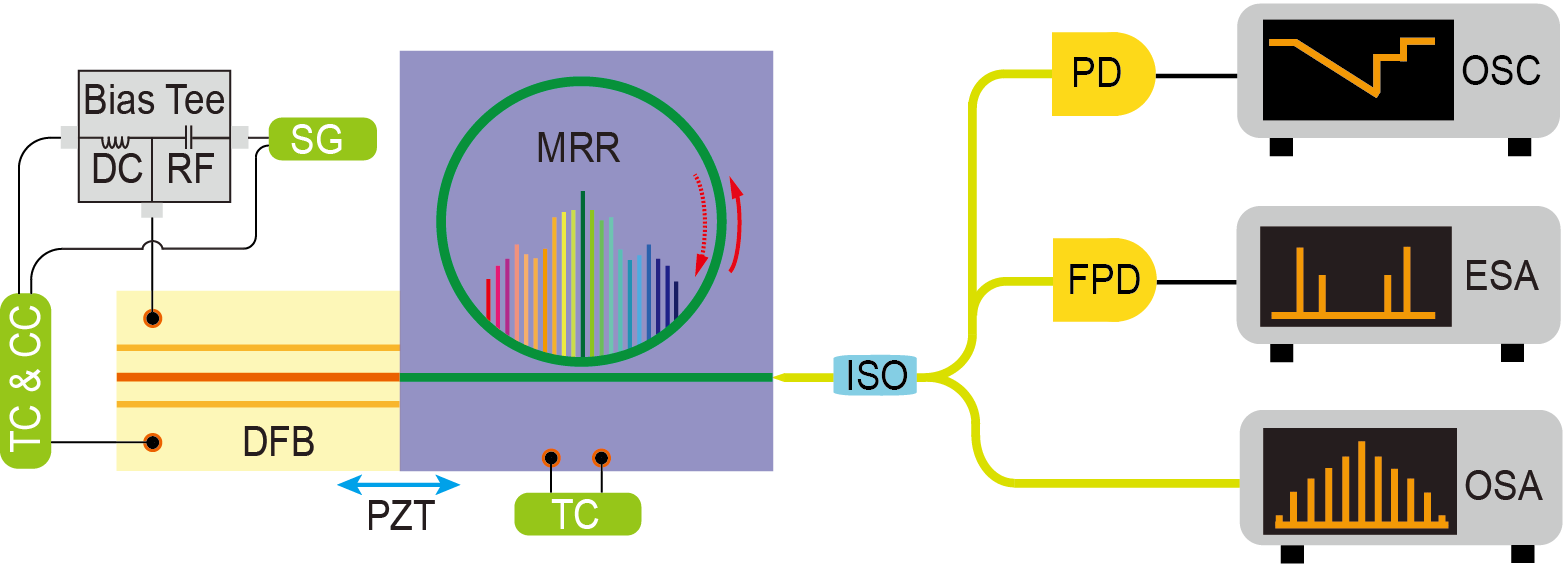}
\caption{The sketch of the experimental setup. TC and CC - temperature and current controllers; DFB - distributed feedback laser diode; PZT - piezo controller; TC - temperature control driver; MRR - integrated Si$_3$N$_4$ ring microresonator; ISO - isolator; PD and FPD - slow and fast photodetectors; OSC - oscilloscope; OSA - optical spectrum analyzer; ESA - electrical spectrum analyzer; Bias-Tee - bias-tee circuit; SG - signal generator. 
}
\label{fig:Setup}
\end{figure}

\begin{figure*}[htbp!]
\centering
\includegraphics[width=\linewidth]{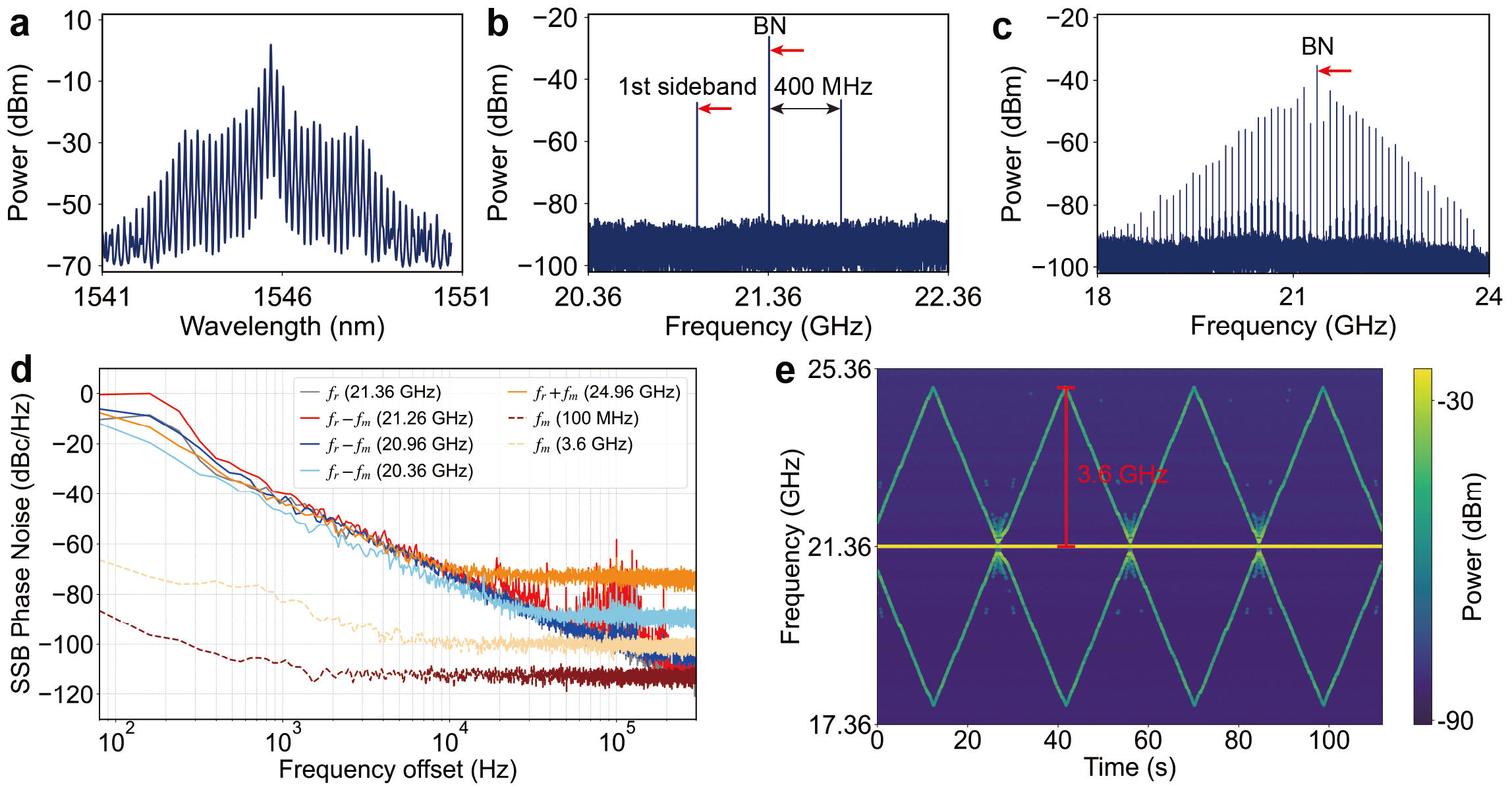}
\caption{(a) Optical spectrum of the GS platicon microcomb. (b) Microwave spectrum of the platicon beatnote generated with the GS laser with $f_m$ = 400 MHz. (c) Regime of wide microwave comb. (d) The SSB phase noise for the first sideband signals for the modulation frequency of $f_m$ = 100 MHz (red), $f_m$ = 400 MHz (blue), $f_m$ = 1 GHz (teal) and $f_m$ = 3.6 GHz (yellow) compared to a platicon beatnote phase noise without GS (grey). The phase noises are almost equal. The beatnotes between the GS optical components at $f_m$ = 0.1 and 3.6 GHz are presented with dashed lines in burgundy and beige colors. (e) More than 100 s spectrograms for the beatnote of platicon in GS regime and its sideband lines for a linear frequency sweep from 100 MHz to 3.6 GHz using ramp modulation.
}
\label{fig:SIL_frequency}
\end{figure*}

\subsection{Experimental setup}

The experimental setup is shown in Fig.~\ref{fig:Setup}. Silicon nitride MRR is pumped by a butt-coupled laser diode. A DFB diode laser serves as a pump source with single-frequency emission centered near 1546 nm with output power up to 100 mW. The laser is supplied through a bias-tee circuit, which enables a gain-switch modulation. The signal generator determines the GS modulation frequency. The temperatures of both the laser diode and Si$_3$N$_4$ chip are stabilized (TC). The microresonator has Q-factor up to 10 million and normal GVD in nearly critical coupling \cite{sun2025chip}. We used microresonators with FSR of 10.7 and 21.3 GHz \cite{li2025universal}. The transmitted signal is analyzed at oscilloscope (OSC), optical spectrum analyzer (OSA), electrical spectrum analyzer (ESA).


\subsection{The platicon beatnote with tunable sidebands via the gain-switched laser diode}

We generate a platicon in the SIL regime and GS sideband lines simultaneously, providing the platicon beatnote, as well as the beatnotes of the first comb lines with the GS sidebands. In this experiment, we use microresonator with FSR of 21.36 GHz. The optical and electrical spectra of the platicon with the modulation frequency of 400 MHz are presented in Fig.~\ref{fig:SIL_frequency}(a,b). 
It is worth noting that the GS sidebands were detected only around the pump, no any sidebands were resolved around platicon comb lines. 
Also, the GS regime can be turned on and off without disturbing the platicon generation. We are able to initiate a SIL-GS laser regime with dozens of components in the range wider than 10 GHz near the pump, see Fig.~\ref{fig:SIL_frequency}(c). The realization of such regimes depends on many parameters that affect the laser diode refractive index, such as the operating current and the locking phase. Such wide GS combs can be extremely useful in spectroscopy, as they provide a broad, dense frequency spectrum with a known envelope in the optical range, which is then extended to the microwave range by the Kerr comb.
To characterize the stability and purity of these signals, we measure the in-phase and quadrature components (I/Q data) of the sideband signals to extract SSB density of the phase noise. All sideband signals exhibit SSB phase noise lower than -70 dBc/Hz at 10 kHz frequency offset (see solid lines in Fig.~\ref{fig:SIL_frequency}(d)). We tested different modulation frequencies from 100 MHz to 3.6 GHz and show that the phase noise of those is almost equal (Fig.~\ref{fig:SIL_frequency}(d)). The only difference can be seen for the frequency offsets above 10$^4$ Hz where high frequency modulation demonstrates higher shot noise, since its optical power is lower. The high coherence of the obtained components can be explained in the following way. The microwave sidebands are the result of the interaction of the nearest to the pump lines of the Kerr comb with the sidebands of the GS pump. The platicon teeth are coherent to each other, including the pump. The GS components are mutually coherent and certainly coherent with the pump. Consequently, since both the platicon and GS components are coherent to the pump, they are mutually coherent. The phase noise of the platicon beatnote in our case is higher than the GS beatnote, so it determines the phase noise of the generated sidebands. The beatnotes between the lines of GS at $f_m$ = 0.1 and 3.6 GHz are presented in Fig.~\ref{fig:SIL_frequency}(d) with dashed lines (burgundy and beige colors consequently). 

Next, we perform an automated frequency sweep in the range from 100 MHz to 3.6 GHz and record the beatnote of the microcomb and its sideband lines to show the tunability of the generated sidebands. We sweep linearly and sinusoidally back and forth the modulation frequency (Fig. \ref{fig:SIL_frequency}(d, e)). We perform multiple cycles to exhibit the remarkable stability of the regime.  
A higher frequency sweep of the GS comb can be obtained by using a signal generator with higher maximum frequency generation. 

So that, we obtain a tool to synthesize any microwave frequency around platicon repetition rate with the same spectral purity. It is important that the phase noise of the sidebands is equal to the phase noise of the platicon beatnote at any applied modulation frequency.

\subsection{Modulation at a frequency whose harmonics is equal to microresonator FSR}

\begin{figure}[hbtp!]
\centering
\includegraphics[width=\linewidth]{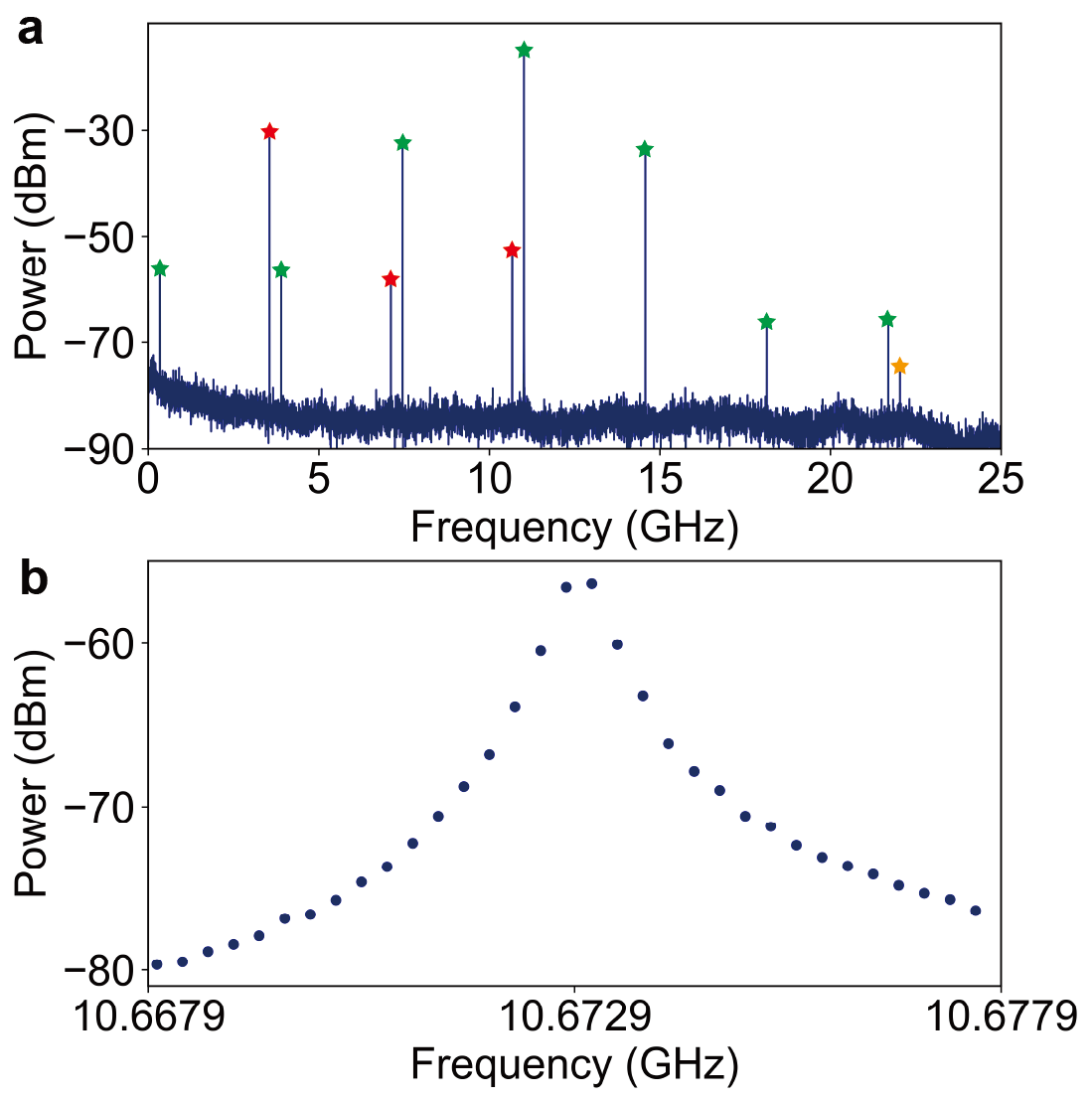}
\caption{(a) Microwave spectrum of the GS laser which is modulated at a frequency fractional to FSR of the microresonator. The red stars mark the interline beatnotes, the green stars mark GS-laser output components heterodyned with reference laser. (b) The dependence of the electrical power of $3f_m$ on its frequency in the vicinity of the microresonator mode. 
}
\label{fig:third_harm}
\end{figure}

Previously we consider only the first sideband of the SIL GS comb by applying just enough modulation voltage for this. Increasing the modulation voltage we can get several more pronounced GS components. In this experiment, we use microresonator with 10.674 GHz FSR. We select the modulation frequency so that the one of the GS harmonics coincides with the FSR of the available MRR.
We use the lower laser output power to avoid frequency comb generation staying in the linear regime. In Fig.~\ref{fig:third_harm}(a) the microwave spectrum of the GS-laser heterodyned with the reference laser in the case of matching third sidebands of the GS-laser with the neighboring microresonator modes is presented. The red stars mark the interline beatnotes, and the green stars mark GS-laser output components heterodyned with reference laser. The third component is generated efficiently and appeared to be more powerful than the second one. It is worth noting that not only neighboring to the pump modes of the microresonator become initiated but also the others through thresholdless four-wave mixing. 
When the triple $f_m$ approaches FSR of the microresonator, the power of $3f_m$ component increases for more than 20 dB at exact match. The dependence of the electrical power of $3f_m$ on its frequency in the vicinity of the microresonator mode is presented in Fig.~\ref{fig:third_harm}(b). This increase in power is a result of the appearance of the backscattering wave from the microresonator at frequencies of neighboring modes. As a result, the third sideband components get optical feedback, which in turn is amplified in the laser. This effect of the sideband amplification can be highly beneficial for gain-switched disciplined platicon combs, which are under consideration in the next section.

\subsection{Microcomb locking via the gain-switching regime}

\begin{figure}[hbtp!]
\centering
\includegraphics[width=\linewidth]{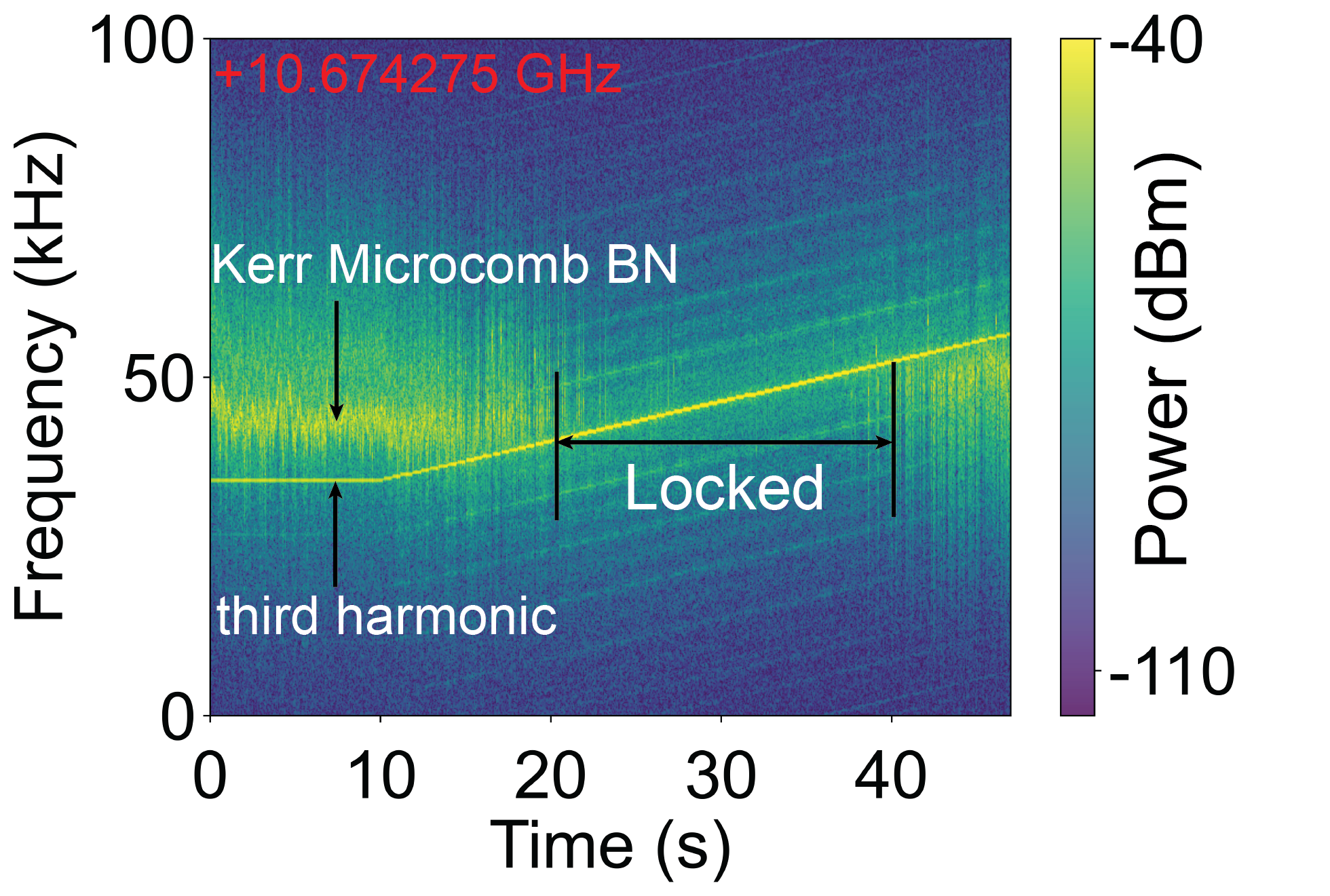}
\caption{The spectrogram of the microwave signal around repetition rate of the platicon when the third-order GS sideband is tuned across $f_r$. In the range of 20 kHz, when $3\cdot f_m = f_r$, the platicon beatnote becomes disciplined, see the area marked as "Locked".
}
\label{fig:spectro}
\end{figure}

\begin{figure*}[hbtp!]
\centering
\includegraphics[width=\linewidth]{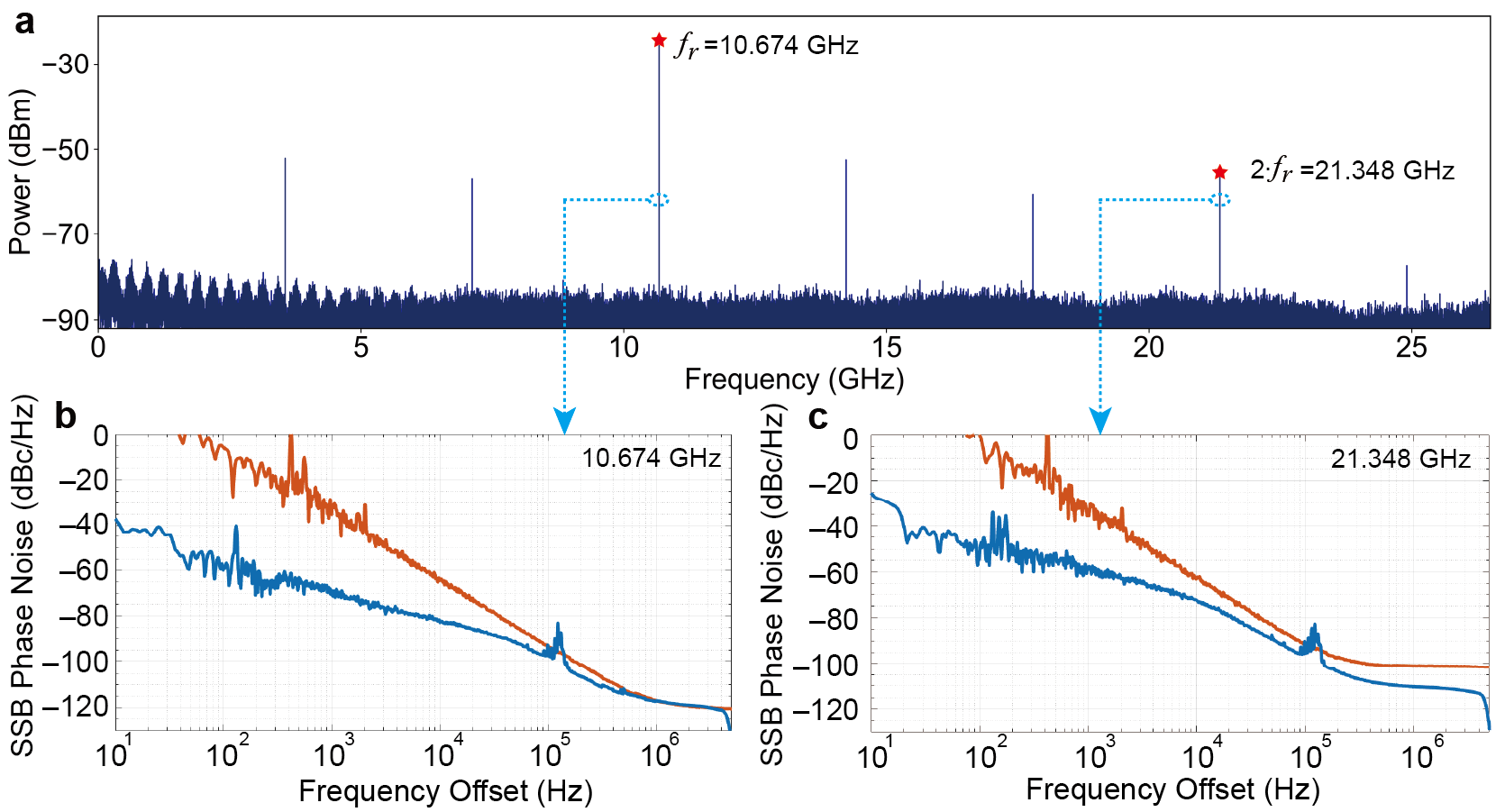}
\caption{(a) Microwave spectrum of the GS locked platicon microcomb. (b,c) Comparison of SSB phase noise spectral densities of the locked and free-running platicon beatnotes at $f_r$ and $2\cdot f_r$. Both components are stabilized at longer measurement time reaching up to 30 dB advancement. 
 }
\label{fig:IL BN}
\end{figure*}

We combine the platicon generation and GS modulation matching the third order GS sideband to first platicon sidebands: $3\cdot f_m = f_r$. First of all, the interaction of the GS sideband with neighboring mode of the microresonator does not interrupt platicon generation. We measure a spectrogram of the frequency scan in the vicinity of the point $3\cdot f_m=f_r$ across the repetition rate frequency (Fig.~\ref{fig:spectro}). Initially we see platicon repetition rate fluctuating and the stable third order GS sideband. Then we start increasing $3\cdot f_m$ linearly towards $f_r$. At some point $f_r$ becomes disciplined by the GS sideband, see the area marked as "Locked" in Fig.~\ref{fig:spectro}. In this area, marked as locked, $f_r$ precisely follows $3\cdot f_m$ in the range of approximately 20 kHz. This behavior can be used for further locking the platicon beatnote to stable ethalon.  
Placing $n\cdot f_m$ in this locked area one may obtain stable generation of GS disciplined platicon. The optical spectrum of the platicon does not noticeably changes in locked regime. The microwave spectrum is presented in Fig.~\ref{fig:IL BN}(a). The platicon beatnote $f_r$ and its second harmonic are marked with red stars. It is worth noting that actually there are GS components at $f_r$ and at $2\cdot f_r$, but they all exactly coincide with the platicon beatnote. We compare the SSB phase noise spectral density of the $f_r$ and $2\cdot f_r$ for the locked and free-running regimes. For frequencies above 10$^3$ Hz, there is no much difference between signals. For frequencies lower than 10$^3$ Hz, the phase noise of signals in the locked regime demonstrates a dramatic improvement up to 30 dB. The obtained values of the phase noise for frequencies below 10$^3$ Hz are equal to the phase noise of the third-order GS sideband. So that, we demonstrate that using 3.557 GHz generator we may synthesize a frequency of 10.67 GHz through GS disciplined platicon along with multiple frequencies as $2\cdot f_r$ = 21.34 GHz. We also observed a purification using forth and fifth order harmonics ($f_m=f_r/4\ or\ f_r/5$) but it was not that efficient since the electrical power of that components was lower.

\section{Discussion}

One can obtain any frequency $n \cdot f_r \pm k \cdot f_m$ where the first term is limited by the bandwidth of the used photodetector, second term is limited by the maximum switching rate of the laser diode, and $f_m$ can be tunable. Unnecessary spectrum components could be filtered out easily.
Such an approach allows one to obtain agile microwave synthesis in a very compact device until one has a photodiode with a high enough bandwidth. Such photodiodes can be placed directly on the chip \cite{li2025sub, sun2025chip}. 

Another interesting possibility is to use the reported tunable GS comb for spectroscopy. We show that it can cover more than 10 GHz spectral range while tuned. Additionally, the platicon lines act as heterodyne, providing a downconversion of the optical comb into the microwave range. In this way, the resulting signal can be analogs of dual-comb spectroscopy \cite{jerez2016dual, quevedo2020gain, coddington2016dual, millot2016frequency}.  Also, wide GS comb regimes demonstrated in Fig.~\ref{fig:SIL_frequency} pave the way to straightforward observation of the spectral lines.

The effect of the purification of platicon beatnote has two straightforward applications. The first one is based on the increased stability at long measurement time. It allows for better resolution in dual-comb spectroscopy. It provides better performance for coherent communications. The second application can be applied for further stabilization of the repetition rate of the platicon. The area where platicon repetition rate becomes locked has a linear dependence on the tripled modulation frequency, providing a convenient way for repetition rate manipulation. It can be locked, for example, to the hyperfine transition in $^{87}$Rb at 6.834 GHz providing exceptional long-term stability to the comb repetition rate and its multiples. 


\section{Conclusion}
In conclusion, we propose a novel approach for Kerr comb generation with a gain-switched laser diode in the self-injection locking regime. This scheme is simple and efficient for purifying all Kerr comb lines using a low-noise external microwave oscillator at the FSR subharmonic. The stability of the GS microcomb beatnote is transferred to the stability of the Kerr microcomb beatnotes. Furthermore, gain-switching regime of the laser provides easily tunable sidebands with low phase noise around platicon beatnotes. Both realizations are attractive for applications that requires high‑quality microwave signals, such as precision timing, spectroscopy, and high‑speed communications.


\bibliography{sample}

\end{document}